\documentclass[letter]{aa} 

\usepackage{graphicx}

\usepackage{txfonts}
\newcommand{\teff}{$T_{\rm eff}$ }

\newcommand{\umpf}{\object{2MASS~J18082002$-$5104378}}
\newcommand{\ump}{\object{2MASS~J1808$-$5104}}
\newcommand{\abund}[2]{\ensuremath{[\mathrm{#1}/\mathrm{#2}]}}

\newcommand{\metal}{\abund{Fe}{H}}

\begin{document}

   \title{2MASS~J18082002$-$5104378: The brightest (V=11.9) ultra metal-poor star\thanks{Based on observations obtained at the
European Southern Observatory (ESO) Very Large Telescope (VLT, observing program 293.D-5036)
and New Technology Telescope (NTT, observing programs 091.D-0292 and 092.D-0308).}
}

\titlerunning{The brightest ultra metal-poor star}

      \author{Jorge Mel\'{e}ndez\inst{1}
          \and 
          Vinicius M.  Placco\inst{2}
          \and
          Marcelo Tucci-Maia\inst{1}
          \and
          Iv\'an Ram{\'{\i}}rez\inst{3}
          \and
          Ting S. Li\inst{4}
          \and
        Gabriel Perez\inst{5}
          }

   \institute{Universidade de S\~ao Paulo, IAG, Departamento de Astronomia,  Rua do Mat\~ao 1226, 
              Cidade Universit\'aria, 05508-900 S\~ao Paulo, SP, Brazil. \email{jorge.melendez@iag.usp.br}
         \and
             University of Notre Dame, Department of Physics and JINA Center for the Evolution of the Elements, 
                         Notre Dame, IN 46556, USA
         \and
         University of Texas at Austin, McDonald Observatory and Department of Astronomy, USA
         \and
         Texas A\&M University, Department of Physics and Astronomy, USA
        \and 
        Universidade de S\~ao Paulo, IAG, Departamento de Ci\^encias Atmosf\'ericas, Rua do Mat\~ao 1226, 
              Cidade Universit\'aria, 05508-900 S\~ao Paulo, SP, Brazil.}

   \date{Received ... 2015; accepted ... 2015}

 
  \abstract
   {The most primitive metal-poor stars are important for studying the conditions of
   the early galaxy and are also relevant to big bang nucleosynthesis.}
   {Our objective is to find the brightest (V $< 14$) most metal-poor stars.}
   {Candidates were selected using a new method, which is based on the mismatch between spectral types derived from 
   colors and observed spectral types. They were
   observed first at low resolution with EFOSC2 at the NTT to obtain an initial
   set of stellar parameters. The most promising candidate, \umpf{} (V = 11.9),
   was observed at high resolution (R = 50 000) with UVES at the VLT, and a standard
   abundance analysis was performed.}
   {We found that \umpf{} is an ultra metal-poor star with stellar parameters
   \teff = 5440 K, log $g$ = 3.0 dex, v$_t$ = 1.5 km s$^{-1}$, [Fe/H] =
   $-$4.1 dex.  The star has [C/Fe] $<$ +0.9 in a 1D analysis, or [C/Fe] $\lesssim$+0.5 
   if 3D effects are considered; its abundance pattern is typical of normal (non-CEMP)
   ultra metal-poor stars. Interestingly, the star has a binary companion.}
   {\ump{} is the brightest (V =  11.9) metal-poor star of its category, and
 it   could be studied further with even higher S/N spectroscopy to
   determine additional chemical abundances, thus
providing    important constraints to the early chemical evolution of our Galaxy.}


   \maketitle
%

\section{Introduction}

The quest for the most metal-poor stars is one of the most active fields in
astronomy. Efforts to discover metal-deficient
stars have been in place for several decades.  Previous surveys
have found a number of them, which are important for
studying the early stages of our Galaxy
\citep[e.g.,][]{bon80,bes84,bee85,bee92,coe04,fre05,
ful10,pla11,caf11,aok13,yon13,kel14,fre15}.

The ultra metal-poor stars ([Fe/H] $< -4$, UMP) are key to studying the products
ejected by the first massive stars (Population III) that polluted the early
Galaxy \citep[e.g.,][]{fre15}. At such low metallicities, the spectral lines are
weak, therefore it is highly desirable to find relatively bright metal-poor
stars, so that detailed abundance analyses can be performed.  

Most of the previous and ongoing surveys focus on finding new faint metal-poor
stars, but their faintness means that they are harder to observe at high spectral
resolution. Nevertheless, some bright extremely metal-poor stars remain to be
discovered, as shown by the bright (V = 9.1) star BD$+$44$^{\circ}$493, identified by
\cite{ito09} as an extremely metal-poor star with [Fe/H] = $-3.7$.  This star is
bright enough not only for high resolution optical spectroscopy but also for
ultraviolet observations with the Hubble Space Telescope, as shown in a recent
work by \cite{pla14}. Based on the lack of molecular absorption in the mid
infrared, another recent work has identified a large number of bright (V $<$
14) metal-poor stars \citep{sch14}.

We have developed a new method that aims to find bright metal-poor stars with V
$<$ 14, in particular those with V $<$ 12, for which high S/N high
resolution spectra can be obtained in large telescopes. In this letter we report
the first bright UMP star found in our survey, \umpf{} (hereafter \ump), a
11.9-magnitude star with [Fe/H] = $-4.07$.

\section{Sample selection and observations}

Our metal-poor candidates were found using an ingenious new method. Metal-poor
stars are usually misclassified in most spectral classification surveys, such as in the case of the two stars (HD~19445 and HD~140283) for which a significant metal deficiency
was discovered for the first time by \cite{cha51}. As noticed by the authors, these
two stars are classified as A-type stars, but actually they were much cooler than stars of spectral
type A. The reason for this mismatch was their low metal abundance,
currently known to be [Fe/H] $\sim -2$ \citep[e.g.,][]{kor03,zha05,nis07,ram13}.

We quantified the difference
between the expected spectral type based on effective temperatures and their
actual spectral classification using a sample of about 7,000 stars for which
effective temperatures (T$_{\rm eff}$), metallicities, and spectral classification are available in the
updated catalog of stellar parameters described in \cite{rm05a}. 
Each spectral subclass was assigned a numeric spectral type ST: ST = 0 for M9, ST = 9 for a M0 star, and so on (ST = 10 for K9, etc.). 
For example, the difference between F0 (ST=39) and G1 (ST=28) is $\Delta$ST = 11.
We calibrated ST with T$_{\rm eff}$ for dwarf (ST = 52.65 ln (T$_{\rm eff}$) - 429.0) and 
giant (ST = 42.10 ln (T$_{\rm eff}$) - 336.4) stars, so that 
ST can be obtained based on T$_{\rm eff}$.
As expected from their
effective temperatures, we find that the most metal-poor stars have the largest difference ($\Delta$ST) 
between their observed spectral types. 

We applied our technique to three large catalogs of spectral classifications
\citep{nes95,wri03,ski10}. We cross-matched those three catalogs to
APASS\footnote{https://www.aavso.org/apass} and 2MASS photometry \citep{cut03},
finding 40055, 195495, and 145441 matches, respectively. Using that photometry
we estimated effective temperatures using color-\teff calibrations
\citep{rm05b,cas10}. We then selected
stars with the largest differences ($\Delta$ST $> 10$) as the most promising metal-poor candidates.  About two
thousand metal-poor candidates were selected for follow-up medium-resolution spectroscopy.
Most of them have 8.5 $<$ V $<$ 14.0, which is therefore adequate for an initial characterization 
of their stellar parameters with medium-resolution spectroscopy at small and medium-size
telescopes and for follow up of the most interesting candidates using high S/N, high-resolution spectroscopy
at medium-size and large telescopes.

Most medium-resolution spectra were acquired with the EFOSC2 spectrograph
(grism \#7) on the NTT telescope at the ESO La Silla Observatory.  Additional
medium-resolution spectra have been acquired 
with the 1.6-m telescope at the Observatorio Pico dos Dias in Brazil and 
with the 2.1-m Otto Struve telescope at the McDonald Observatory in the USA.  The spectra were reduced
with IRAF, following standard procedures (bias subtraction, flat fielding, sky
subtraction, extraction of the spectrum, wavelength calibration).

\begin{figure}
\resizebox{\hsize}{!}{\includegraphics{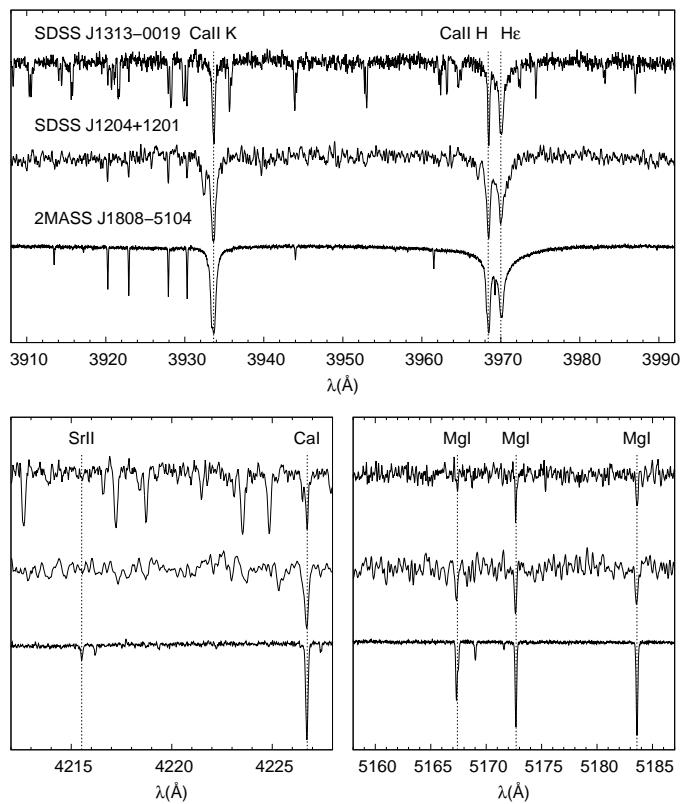}}
\caption{Comparison of 2MASS J1808-5104 to other UMP stars \citep{pla15,fre15}.
Both SDSS stars were observed with Magellan/MIKE, at a resolution of R$\sim$35,000.
Our UMP star was observed at R = 50,000 and has a higher S/N ratio than SDSS stars.}
\label{spec_comp}
\end{figure}

The stellar parameters of the medium-resolution spectra were estimated using a
modified version of the SEGUE Stellar Parameter Pipeline \cite[e.g.,][]{bee14}.
The analysis revealed that \ump{} was probably a subgiant with [Fe/H] $\sim -4$,
so we selected this star for high-resolution follow-up
spectroscopy. Observations were acquired with the UVES spectrograph at the VLT
telescope (program 293.D-5036) at a resolving power R = 50,000.  We used the
dichroic \#1 with the cross disperser \#2 in the blue (330-450nm) and the cross
disperser \#3 in the red (480-680nm).  Three exposures of 1800 s were obtained
adding to a total exposure time of 1h30m. The high-resolution spectra were
reduced using the UVES pipeline, and Doppler correction and continuum
normalization were performed with IRAF. Figure~\ref{spec_comp} shows portions of
the observed spectrum, compared with SDSS~J1204$+$1201 \citep[\metal = $-4.34$,
\teff=5467~K;][]{pla15} and SDSS~J1313$-$0019 \citep[\metal = $-5.00$,
\teff=5200~K;][]{fre15}.  

The measured radial velocities are 70.4, 66.5, and -29.1 km/s ($\sigma$ = 0.6
km/s) in 19 October 2014, 21 October 2014 and 6 March 2015, respectively, showing that
the star has a binary companion. No sign of the secondary is present in the
spectra (whether on the blue or the red side), therefore we combined the
individual spectra to achieve a S/N of about 150 per resolution
element on the blue side and 300 on the red.

\section{Analysis}

The techniques applied to determine the stellar atmospheric parameters are
described in detail in \cite{pla15b}.  Most of the work is based on equivalent
widths (EW) measured by fitting Gaussian profiles to the observed atomic lines,
using the {\tt Robospect} package \citep{robo}. The line lists used within {\tt
Robospect} were compiled from \citet{roe12}, the VALD database \citep{vald}, and
the National Institute of Standards and Technology Atomic Spectra Database
\citep[NIST; ][]{nist}.  The model atmospheres are one-dimensional
plane-parallel, no overshooting models computed in LTE from the grid of
\cite{cas04}.

For the stellar atmospheric parameter determination, as well as abundances from
EW analysis and spectral synthesis, we used the latest version (2014) of the MOOG
analysis package \citep{sne73}. 
The effective temperature was determined by imposing the excitation equilibrium of
Fe I lines, and then we used the corrections described in \citet{fre13} to bring 
the spectroscopic temperature into agreement with the photometric scale.
The surface gravity was obtained by ionization equilibrium of Fe I and Fe II, and
the microturbulence was estimated by enforcing no trend of Fe I abundances with
reduced equivalent width.
Our stellar parameters are \teff = 5440$\pm100$~K, log$g$ =
3.0$\pm0.2$~dex, v$_t$ = 1.5$\pm0.2$~km~s$^{-1}$, and [Fe/H] = $-$4.07$\pm0.07$~dex.  

Chemical abundances were derived from EW analysis for Na, Mg, Al, Si, Ca, Sc,
Ti, Cr, Mn, Fe, Co, and Ni. From spectral synthesis, we were able to measure
abundances for Li and Sr, and upper limits for C, N, and Ba. For Ba, the upper
limit was determined from the 4934\AA\ feature, since the strong 4554\AA\ line
fell in one of the gaps of the CCD. The elemental abundances and errors are
shown in Table~\ref{abund}.

\section{Discussion}

\begin{figure}
\resizebox{\hsize}{!}{\includegraphics{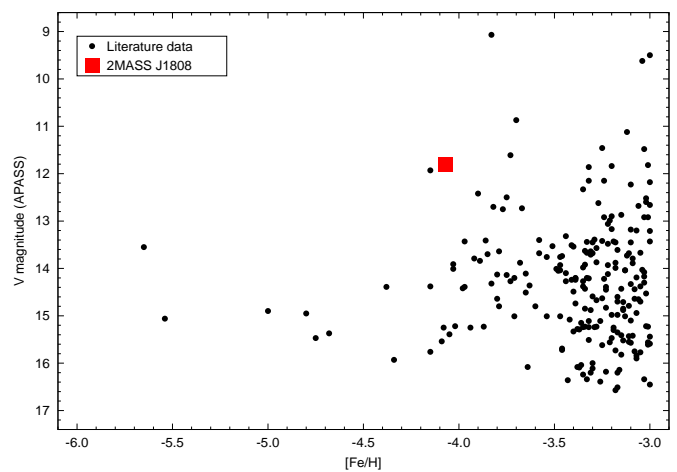}}
\caption{
V magnitudes for metal-poor stars with [Fe/H] $\leq -3$.
\protect{\ump}~ is the brightest UMP ([Fe/H] $<$ -4).
The literature data are from the compilation presented in \cite{pla14b}.
}
\label{vfeh}
\end{figure}

With V = 11.9 and [Fe/H] = $-4.1$, \ump{} is the brightest UMP star (see
Figure~\ref{vfeh}).  The star is slightly brighter than the giant UMP star
CD$-$38~245 (V = 12.0), which has stellar parameters \teff = 4857~K, log$g$ =
1.54, and [Fe/H] = $-4.15$~dex \citep{yon13}.

The carbon abundance in a 1D analysis is [C/Fe] $<$ +0.94 or +0.87, depending on the adopted
solar carbon abundance \citep{asp09,caf10}. 
However, the CH lines are affected by strong 3D effects at low metallicity \citep[e.g.,][]{bon09}.
Although there are no predictions regarding
the 3D effects for a star like \ump{}, 
interpolating on calculations for stars in the range of \teff = 4880 - 6550 K, log $g$ = 2.0 - 4.5, and
$-5 <$ [Fe/H] $< -3$ \citep{bon09,caf12,spi13}, the correction would be $\Delta$C(3D - 1D) $\sim -$0.4 dex, meaning
that in 3D the carbon abundance would be [C/Fe] $\lesssim+0.5$.

According to \cite{aok07}, a carbon-enhanced metal-poor star (CEMP) has
[C/Fe] > +0.7. However, we notice that this suggestion was mostly based on
stars with [Fe/H] $> -3.5$, and as the [C/Fe] ratio seems to increase
at lower metallicities, perhaps this definition should be revised.
\cite{bee05} define CEMP stars as those with [C/Fe] > +1.0, a definition
also adopted by \cite{spi13}. If the normal (non-CEMP) stars indeed have
higher [C/Fe] at lower metallicities \citep[see, e.g., Fig. 13 in][]{pla14b}, perhaps a 
metallicity-dependent definition of CEMP should be considered.
In any case, the 3D-corrected value of [C/Fe] $\lesssim$+0.5, surely puts
\ump{} in the category of non-CEMP star.

At the metallicity and log $g$ (= 3.0$\pm$0.2) of our UMP star, it should be either leaving 
the subgiant branch or entering the bottom of the red giant branch (RGB), 
because it is less evolved than the giant (log $g$ = 1.5) CD$-$38~245. This is important for
the study of the light elements (C, N, O), since \ump{} should be much less affected by mixing
than CD$-$38~245.  
According to observations and calculations (including extra-mixing)
for field and globular cluster giants \citep{gra00,den03,den15,sta09,she10,ang11},
during the first dredge-up (starting at M$_V$ $\sim$ 2.0, log $g \sim$3.0, log L/L$_\odot \sim$ 1.2),
there should only be a minor depletion of carbon ($\Delta C < 0.1$ dex). 
Only when the star reaches the RGB bump (M$_V$ $\sim$ $-$0.2, log $g \sim$2.0,  log L/L$_\odot \sim$ 2.1)
does a steep carbon depletion begin ($\Delta C >> 0.1$ dex).

Because CD$-$38~245 is above the RGB bump, it should have already experienced 
deep mixing. Indeed, only an upper limit is available for its carbon
abundance  \citep[see, e.g.,][]{spi05}. This upper limit in C
and its N abundance place CD$-$38~245 among the mixed stars of the sample of
metal-poor giants of \cite{spi05}. With higher S/N data, we can try to measure
the C and N abundances in \ump{}, so as to have a measurement of the abundances of
those elements in the primeval Galaxy.

\begin{figure*}
\resizebox{\hsize}{!}{\includegraphics{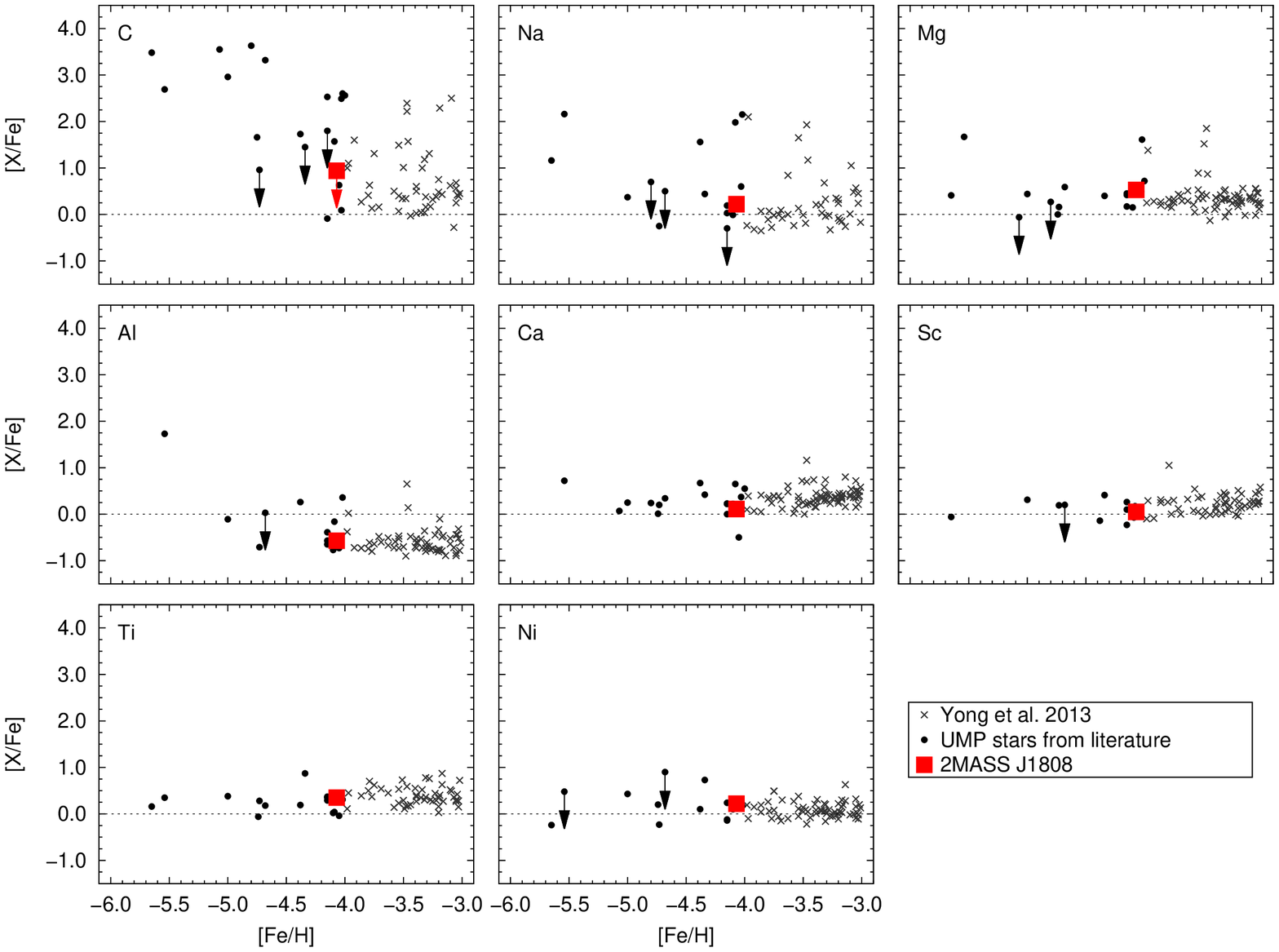}}
\caption{Chemical abundances of \protect{\ump}~
compared to metal-poor stars with [Fe/H] $\leq -3$ from \cite{yon13}
and UMP from the compilation by \citet{pla15}.
}
\label{abfullh}
\end{figure*}

In Figure~\ref{abfullh} the abundance pattern of \ump{} is compared to
the sample of \cite{yon13} and UMP stars from the literature, as compiled by
\citet{pla15}. The [X/Fe] ratios match other UMP stars, and also fits the 
extension of the \cite{yon13} pattern to the UMP regime.

The lithium abundance, $A$(Li) = 1.45~dex, places \ump{} in the
group of ``unmixed'' (meaning without deep mixing) stars of \cite{spi05}, thus confirming that \ump{}
is key to studying the light elements, since its abundances should reflect the ``pristine''
values of the time that the star was formed.
For the neutron-capture elements, we were only able to determine an abundance
for Sr, $A$(Sr)=$-2.03$, and an upper limit for Ba, $A$(Ba)$<-2.20$. These
values are consistent with the ``abundance floors'' suggested by \citet{han15}
for UMP stars.

As a non-CEMP, very metal-deficient star ([Fe/H] = $-$4.1), 
\ump{} is important for understanding the percentage of CEMP stars at low metallicity
\citep{pla14b} and the channels that can form CEMP-no stars at
such extreme metallicities \citep[e.g.,][]{coo14}. Perhaps this could be
related to the binarity of \ump{}, because population III stars have a channel
of low-mass binary formation \citep{sta14}.

Although \ump{} is a binary, the light elements are not enhanced. This
star has an abundance pattern typical of other non-CEMP stars.
This is consistent with mass transfer without significant enhancement of
the light elements, according to binary population synthesis \citep{sud13}.

\section{Conclusions}
   
      Our new technique, which is based on the mismatch 
      between the expected spectral type derived from the effective temperature of a star and its actual spectral classification, 
      can successfully identify metal-poor stars.
We have identified \ump{} as the brightest (V = 11.9) UMP star ([Fe/H] $< -$4).
      
Variations in the radial velocities of this star, show that it belongs to a binary system, 
      but the spectrum of the secondary was not detected. 
      The star has [C/Fe] $\lesssim$+0.9 in a 1D analysis or [C/Fe] $\lesssim$+0.5 if 3D effects are considered,
      making it a non-CEMP star. The abundance pattern is also typical of non-CEMP stars.
       The star seems to be leaving the subgiant branch or entering the bottom of the
      RGB, making it useful for studying the abundances of the light elements because the star should 
      not have experienced any deep mixing yet.

 We intend to perform higher S/N observations  to detect C (from CH) and nitrogen (from NH).
      Also, we would like to increase our ultraviolet coverage, in order to reach other elements, such
      as oxygen (from OH) and heavy elements. Our new observations will also provide
      valuable information on radial velocities, so that we can characterize the binary system.

\begin{acknowledgements}
J.M. acknowledges support from FAPESP (2012/24392-2, 2010/50930-6) and CNPq
(Bolsa de Produtividade). 
V.M.P. acknowledges partial support for this work from PHY 08-22648; Physics
Frontier Center/{}Joint Institute for Nuclear Astrophysics (JINA) and PHY
14-30152; Physics Frontier Center/JINA Center for the Evolution of the Elements
(JINA-CEE), awarded by the US National Science Foundation.

\end{acknowledgements}

%
%

\begin{table}
\caption{Abundances for individual species} 
\label{abund}
\begin{tabular}{lrrrrrrrrrr} 
\hline
\hline
Species & $A_{\odot}$\,(X) & $A$\,(X) & $\mbox{[X/H]}$ & $\mbox{[X/Fe]}$ & $\sigma$ & $N$ \\
\hline
Li           & 1.05 &     1.60 &      {}  &     {}   &    0.15 &  1 \\ 
C\tablefootmark{*}           & 8.43 &  $<$5.30 & $<-$3.13 & $<+$0.94 &    {}   &  1 \\ 
N            & 7.83 &  $<$4.50 & $<-$3.33 & $<+$0.74 &    {}   &  1 \\ 
\ion{Na}{I}  & 6.24 &     2.39 &  $-$3.85 &  $+$0.22 &    0.04 &  2 \\
\ion{Mg}{I}  & 7.60 &     4.07 &  $-$3.53 &  $+$0.53 &    0.01 &  4 \\
\ion{Al}{I}  & 6.45 &     1.81 &  $-$4.64 &  $-$0.57 &    0.10 &  1 \\
\ion{Si}{I}  & 7.51 &     3.57 &  $-$3.94 &  $+$0.13 &    0.10 &  1 \\
\ion{Ca}{I}  & 6.34 &     2.38 &  $-$3.96 &  $+$0.11 &    0.10 &  1 \\
\ion{Sc}{II} & 3.15 &  $-$0.87 &  $-$4.02 &  $+$0.05 &    0.10 &  1 \\
\ion{Ti}{I}  & 4.95 &     1.44 &  $-$3.51 &  $+$0.56 &    0.11 &  3 \\
\ion{Ti}{II} & 4.95 &     1.23 &  $-$3.72 &  $+$0.35 &    0.07 & 21 \\
\ion{Cr}{I}  & 5.64 &     1.21 &  $-$4.43 &  $-$0.37 &    0.05 &  4 \\
\ion{Mn}{I}  & 5.43 &     0.52 &  $-$4.91 &  $-$0.84 &    0.10 &  1 \\
\ion{Fe}{I}  & 7.50 &     3.43 &  $-$4.07 &  $-$0.00 &    0.07 & 73 \\
\ion{Fe}{II} & 7.50 &     3.43 &  $-$4.07 &  $-$0.00 &    0.02 &  3 \\
\ion{Co}{I}  & 4.99 &     1.71 &  $-$3.28 &  $+$0.79 &    0.04 &  3 \\
\ion{Ni}{I}  & 6.22 &     2.37 &  $-$3.85 &  $+$0.22 &    0.09 & 15 \\
\ion{Sr}{II} & 2.87 &  $-$2.03 &  $-$4.90 &  $-$0.83 &    0.15 &  1 \\ 
\ion{Ba}{II} & 2.18 & $<-$2.20 & $<-$4.38 & $<-$0.31 &    {}   &  1 \\ 

\hline       
\end{tabular}
\tablefoottext{*}{C is from our 1D analysis. In 3D the C abundance from CH lines should be about 0.4 dex smaller.}
\end{table}

\end{document}